# ELT instrumentation for seeing-limited and AO-corrected observations: A comparison

Colin Cunningham & Chris Evans
UK Astronomy Technology Centre

Guy Monnet & Miska Le Louarn
European Southern Observatory

## INTRODUCTION

The next generation of large ground-based optical and infrared telescopes will provide new challenges for designers of astronomical instrumentation. The varied science cases for these extremely large telescopes (ELTs) require a large range of angular resolutions, from near diffraction-limited performance via correction of atmospheric turbulence using adaptive optics (AO), to seeing-limited observations. Moreover, the scientific output of the telescopes must also be optimized with the consideration that, with current technology, AO is relatively ineffective at visible wavelengths, and that atmospheric conditions will often preclude high-performance AO. This paper explores some of the issues that arise when designing ELT instrumentation that operates across a range of angular-resolutions and wavelengths. We show that instruments designed for seeing-limited or seeing-enhanced observations have particular challenges in terms of size and mass, while diffraction-limited instruments are not as straightforward as might be imagined.

## THREE INSTRUMENTAL REGIMES

Simulated images of a 'toy' star cluster as observed by a 42-m telescope in K-band are shown in Figure 1. The three images show examples of different observational modes of the telescope: seeing-limited observations, "enhanced seeing", via the use of ground-layer AO (GLAO), and diffraction limited, via laser-tomography AO (LTAO). In this paper we consider three distinct regimes in which ELT instruments will be required to operate:

- Diffraction limited, via XAO (extreme AO), LTAO, or perhaps MCAO (multi-conjugate AO).
- Aperture-ratio (f/#) limited, via MCAO or MOAO (multi-object AO).
- Seeing-limited or GLAO observations.

We now discuss each of these in more depth. For further details of the different AO modes in the context of ELTs, see Hubin et al. [ref. 1].

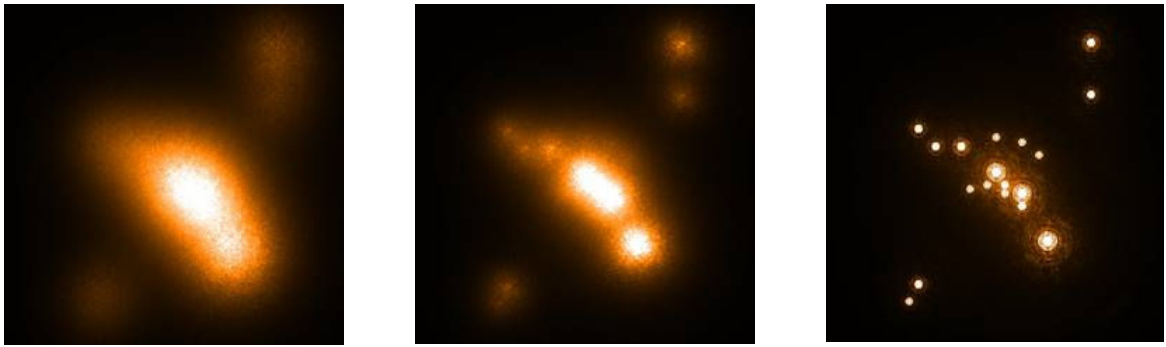

**Figure 1: Simulated E-ELT 2x2 arcsec images generated using SPECSIM [2] to illustrate seeing-limited (left-hand panel), GLAO-corrected (centre), and LTAO-corrected (right) observations.**

## THE DIFFRACTION-LIMITED CASE

When working in the diffraction-limited regime, the étendue (AΩ) is equal to $\lambda^2$. One might therefore assume that such instruments would be scale-invariant with respect to the telescope aperture, and that clones of VLT instruments could satisfy the requirements for science cases that call for diffraction-limited performance. One can imagine a clone of CONICA [3] for diffraction-limited imaging of crowded stellar populations, or a copy of SPHERE [4] for high-contrast imaging and spectropolarimetry of exo-planets. However, we have greater expectations of the science from the ELTs!

For instance, construction of a useful colour-magnitude diagram from observations of crowded stellar populations requires accurate (~1%) relative two-band photometry (e.g., *I* and *K*) over a field of at least ~15", so as to be possible in a realistic amount of time. An ELT clone of CONICA would only have a field-of-view of a few arcseconds, with insufficient photometric accuracy – for the ELT version this suggests a wider-field AO system is needed (e.g. MCAO), combined with more stringent stability requirements. Similarly, the direct detection and characterization of exo-planets via high-contrast, diffraction-limited imaging will be even more demanding than for SPHERE. The increased photon rate from the central star will be $\propto D^2$, i.e. a 42-m aperture will deliver some 25 times the flux from an 8-m primary. This means the required contrast factor scales by a factor of five, putting even more extreme requirements on stability, metrology, and internal calibrations for an ELT instrument such as EPICS [5]; i.e. we cannot just adopt diffraction-limited designs from VLT instruments.

## THE INTERMEDIATE CASE

Many of the high-priority ELT science cases require improved image quality from AO, but can be satisfied by spatial sampling that is well above the diffraction limit. This is the intermediate case, in which the on-sky sampling is bound by the largest étendue of a detector pixel. Current optical and near-infrared detectors have maximum pixel sizes of ~18 μm, and the fastest, practical camera designs (limited by both the optical performance and the flatness of the detector arrays) are ~f/1.75. If we consider an instrument at the f/16 Nasmyth focus of the 42-m E-ELT, one pixel will correspond to (16/1.75) x 18μm = 165 μm in the focal plane. The plate scale at the f/16 focus is 3.25 mm/arcsec, so one pixel corresponds to 51 mas sampling on the sky, which is well-matched to science cases such as near-infrared, spatially-resolved observations of high redshift galaxies. However, before considering instruments to operate in this regime, we compare the performance of different AO concepts to determine which is the most appropriate.

## E-ELT SIMULATIONS: GLAO & LTAO PERFORMANCES

We now present results from simulated GLAO and LTAO point spread functions (PSFs) for the E-ELT. Figure 2 shows the improvement in encircled energy (EE), as a function of the diameter of the PSF, that GLAO provides compared to seeing-limited observations. As one would expect, the correction is most effective at longer wavelengths.

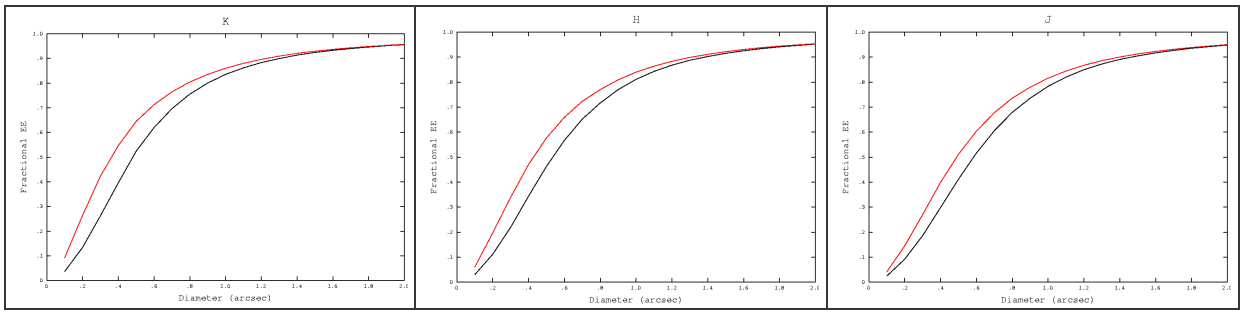

**Figure 2: Fractional encircled energies from simulated K, H and J band PSFs for GLAO (red, upper lines; grey in the printed version) and seeing-limited observations, assuming 0.8" DIMM seeing.**

For a more quantitative comparison, Table 1 summarizes the PSF diameters for 50% EE for the GLAO and seeing-limited PSFs. The simulations assumed a 42-m primary aperture, with an adaptive M4 mirror (with 85x85 actuators, and 84x84 sub-apertures), a Paranal-like turbulence profile with an outer-scale of 25 m, and a seeing of 0.8" – as defined by a Differential Image Motion Monitor (DIMM) telescope at 0.5µm. To investigate the dependence of the GLAO correction on the distance of the laser guide stars (LGS), two batches of simulations were calculated. The first assumed a ring of five LGS at a radius of 3' from the centre of the field, with the LGS at a radius of 3.75' in the second batch – as can be seen from Table 1, the GLAO performance is less effective when the LGS are further away, but only marginally so at these distances. These numbers are also of use in consideration of the physical size that target "pick-offs" (be they mirrors, slitlets, fibres etc.) need to be in the telescope focal plane.

**Table 1: Diameters of 50% EE (in mas) for simulated E-ELT PSFs with 5×LGS at r = 3' and 3.75'. For comparison we also include the diffraction limit at each wavelength (λ/D), the seeing-limited diameters ('No AO'), and results for 50 and 30% EE from LTAO simulations. The DIMM seeing is 0.8" in each.**

| Band | Wavelength [µm] | λ/D | No AO | GLAO LGS @3' | GLAO LGS @3.75' | LTAO [50% EE] | LTAO [30% EE] |
|---|---|---|---|---|---|---|---|
| K | 2.20 | 10.8 | 480 | 360 | 390 | 30 | 10 |
| H | 1.65 | 8.1 | 540 | 430 | 450 | 70 | 20 |
| J | 1.25 | 6.1 | 590 | 490 | 510 | 240 | 40 |
| I | 0.90 | 4.4 | 650 | 570 | 580 | 410 | 190 |
| R | 0.70 | 3.4 | 690 | 620 | 630 | 490 | 280 |
| V | 0.55 | 2.7 | 730 | 670 | 680 | 550 | 350 |
| B | 0.44 | 2.2 | 780 | 720 | 720 | 610 | 400 |
| U | 0.37 | 1.8 | 810 | 750 | 760 | 650 | 440 |

Although GLAO provides a very useful enhancement to the observed PSF, particularly at near-infrared wavelengths, it is clearly not sufficient to match to the spatial sampling discussed above (i.e. 50 mas). In Table 1 we also present results for 50 and 30% EE from LTAO simulations, in which there is a circle of five LGS at a radius of 45", with a sixth in the centre of the field. These are included to highlight the fact that even for the "intermediate case" under discussion, the spatial scales are such that high-performance LTAO is required to achieve a useful encircled energy. If a large multiplex is required, over a field-of-view that is larger than the relatively small patch corrected by LTAO, more sophisticated modes such as MOAO or MCAO will be required. We now mention three example instrument concepts that fall into this intermediate category.

- **E-ELT HARMONI**

    A clone of SINFONI [6] would seem appropriate for observations of one extended object. One of the E-ELT Phase A studies now underway is building on this concept, namely the High Angular-Resolution, Monolithic Optical and Near-infrared Integral field spectrograph (HARMONI). The specifications are still the subject of scientific trade-off, but the spatial scales that will be considered in the design study range from our intermediate case (i.e. 50 mas), down toward critical sampling of the diffraction limit (~5 mas). A high-performance LTAO system is most likely the appropriate system to deliver this spatial sampling (cf. Table 1).

- **E-ELT EAGLE**

    In cases such as resolved spectroscopy of high redshift galaxies (e.g. the right-hand panel of Figure 3), we want to improve on the GLAO performance, but over a much wider field-of-view than delivered by LTAO systems. EAGLE is a two-year, French-UK study of a near-infrared (0.8 to 2.5 µm) instrument that will use MOAO or MCAO to correct small sub-fields within the 5' focal plane of the E-ELT. Deployable integral field units (IFUs) will provide $R$~5,000 spectroscopy in the YZ, J, H, or K bands, of 20 to 60 separate sub-fields. Higher spectral resolution modes are currently under study, primarily driven by the resolved stellar populations science cases.

- **TMT-IRMS**
  The infared multi-object spectrograph (IRMS) is a clone for the Thirty Meter Telescope (TMT) of the MOSFIRE instrument currently under construction for the Keck Observatories [7]. IRMS will be deployed behind the TMT-NFIRAOS unit that will deliver image quality of ~60 milliarcseconds over a 2' field using MCAO [8]. Targets will be selected using a cryogenic slit-mechanism, giving spectra of a full atmospheric band (i.e. YZ, J, H, or K) at *R* of up to 5,000; the configurable slitlets can also be withdrawn to provide a near-infrared imaging capability.

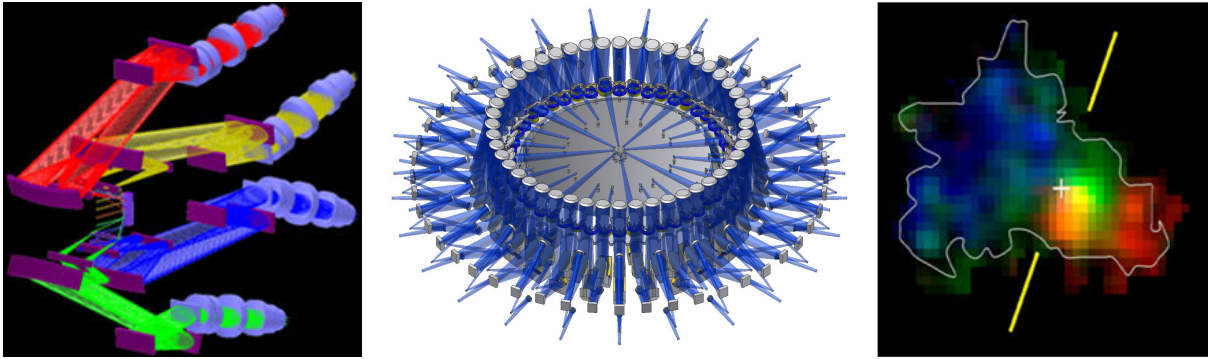

**Figure 3: Early-stage conceptual designs for HARMONI [left; 9] and EAGLE (centre). The right-hand panel shows a rotating disk galaxy at z=2.4, as viewed by SINFONI [10]. A multi-IFU ELT instrument such as EAGLE could complete a large survey of ~1,000 such galaxies in a realistic amount of time.**

As an aside, we note that the LTAO results in Table 1 contain an important result for further development of the ELT science case. The I-band is potentially the most sensitive photometric band – the background is relatively low compared to longer wavelengths, and yet AO still yields a meaningful improvement in image quality. In particular, the calcium triplet (centred at ~0.86 μm) is a powerful diagnostic in studies of stellar populations. However, even in the small-field LTAO case, the effective width of the PSF at 0.9 μm is several tenths of an arcsecond, i.e. **in typical 'Paranal-like' conditions, spectroscopy of the calcium triplet will be limited to spatial resolutions of ~0.2 arcsec at best.** With ongoing technology development it is likely that deformable mirrors with more than 85x85 actuators will be produced, but this is still a decade or more in the future. Remembering that construction of the first instruments will have to commence long before that, capacity for future upgrades of higher-order mirrors should be included in the designs.

## THE SEEING-LIMITED CASE

There is a large and diverse range of potential science targets for which we are 'photon starved' using current facilities. These range from spectroscopic confirmation of the highest-redshift galaxies (which only requires low signal-to-noise, low spectral-resolution), to radial-velocity detection of low-mass exo-planets around nearby bright stars (which demands exceptional signal-to-noise, at high spectral-resolution). Seeing-limited (or seeing-enhanced, via GLAO) observations will benefit immensely from the $D^2$ gain offered by the ELTs. However, the étendue is also proportional to $D^2$, meaning that the sheer physical size of seeing-limited ELT instruments will be somewhat daunting to say the least. One potential solution to this is to split the focal plane between multiple instruments or channels (e.g. Figure 4), each of which would have dimensions comparable to existing 8-m class instruments.

Another problem that arises in the seeing-limited regime is over-sampling of the targets, a consequence of the previous discussion regarding the étendue of a detector pixel. In Table 2 we summarize the results from GLAO simulations in different seeing conditions. As an example, consider J-band observations in median seeing at Paranal of 0.8" [11]. The 50% EE diameter, with GLAO-correction is ~0.5", which corresponds to 1.6mm in the E-ELT focal plane. Making the same assumptions as before (f/16 Nasmyth mount, f/1.75 camera), this is demagnified to 0.175mm at the detector which, assuming 18μm pixels, is 9.7 pixels. This will reduce sensitivity because of the increased detector noise and, with the current cost of infrared arrays, would almost certainly be prohibitively expensive for a large field-of-view.

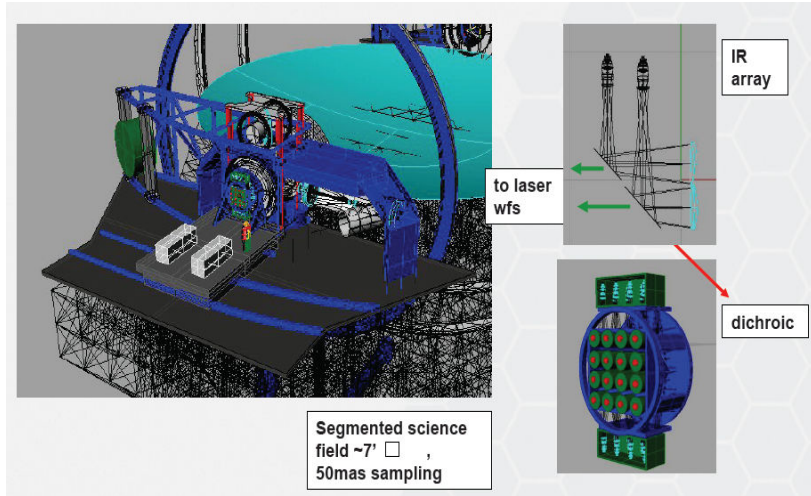

**Figure 4: ESO concept for a wide-field near-infrared GLAO instrument for the E-ELT**

On-chip binning and the relatively low cost of CCD arrays means that over-sampling for optical instrumentation is not such a severe problem. Future developments in near-infrared detectors will likely mean that the noise effects of over-sampling can also be somewhat mitigated. However, while larger pixels are possible (e.g. 25μm pixels in the Orion detectors in the NOAO-NEWFIRM instrument [12,13]), we note that most ongoing detector development is directed toward an increasing the number of smaller pixels on current-sized arrays. Of course, novel technologies might offer different solutions in the future to increase the AΩ-grasp of a pixel. For instance, avalanche photodiode arrays could potentially have ~100μm pixels [14], with the added benefit that the cosmic ray contribution can be reduced owing to the real-time array photon counting capability (although dark noise remains a problem). An alternative solution to the over-sampling problem could be to increase the camera aperture ratio by immersing the detectors in a material with a high index of refraction. However, this is non-trivial for both thinned CCDs and infrared arrays because of their vacuum, cooled/cryogenic environments.

Inclusion of instrumentation optimized for seeing-limited and GLAO operations is also prudent for more prosaic reasons. Even good astronomical sites have very fast, high-altitude turbulence (that AO cannot correct for, apart from removal of the ground-layer profile) some 25% of the time. Availability could also be helped by using natural guide stars (NGS) for the GLAO-correction, yielding similar performances to the LGS simulations reported here, but with reduced sky-coverage – this would have the advantage of being independent of the LGS (less complex operations, engineering down-time, etc.) and also enable observations when cirrus is present (in which use of LGS is difficult or impossible). Another suggestion for sub-optimal conditions is to use them for high time-resolution observations [15].

**Table 2: Diameter of 50% encircled energy for simulated E-ELT seeing-limited ('No AO') and GLAO PSFs with DIMM seeing of 0.4", 0.8", 1.2" and 1.6". Note that all results are quoted in arcsec.**

|      | DIMM = 0.4" | | DIMM = 0.8" | | DIMM = 1.2" | | DIMM = 1.6" | |
|------|-------------|------|-------------|------|-------------|------|-------------|------|
| Band | No AO | GLAO | No AO | GLAO | No AO | GLAO | NoAO | GLAO |
| K | 0.22 | 0.14 | 0.48 | 0.36 | 0.77 | 0.63 | 1.06 | 0.91 |
| H | 0.25 | 0.17 | 0.54 | 0.43 | 0.85 | 0.73 | 1.15 | 1.02 |
| J | 0.28 | 0.20 | 0.59 | 0.49 | 0.92 | 0.81 | 1.24 | 1.12 |
| I | 0.31 | 0.25 | 0.65 | 0.57 | 1.00 | 0.91 | 1.35 | 1.25 |
| R | 0.34 | 0.28 | 0.69 | 0.62 | 1.06 | 0.98 | 1.42 | 1.33 |
| V | 0.36 | 0.31 | 0.73 | 0.67 | 1.12 | 1.04 | 1.49 | 1.40 |
| B | 0.38 | 0.34 | 0.78 | 0.72 | 1.17 | 1.10 | - | - |
| U | 0.40 | 0.36 | 0.81 | 0.75 | - | - | - | - |

# HIGH SPECTRAL-RESOLUTION SPECTROSCOPY IN THE VISIBLE

There are compelling cases for high signal-to-noise (>1,000) spectroscopy with an ELT in the visible, at $R$ = 50,000 to 200,000. We now mention two high-resolution instruments proposed for the E-ELT and TMT:

- **E-ELT CODEX**

  The primary science drivers for the CODEX design study are to constrain the fundamental parameters of the Universe (via direct detection of its acceleration), and detection of low mass exo-planets. Both of these require very accurate radial velocities, with exacting requirements on long-term stability (~2 cms$^{-1}$ over 10 years). To this end, the instrument will be mounted at a fibre-fed Coudé focus, with other factors, such as sensitivity, only optimized to the extent that they do not detract from the stability. The current CODEX design splits the object seeing disk (~1") in the pupil-plane and relays it to five identical cross-dispersed echelle spectrographs (Figure 5), each of which is equivalent in scale to two of the VLT-UVES spectrographs.

- **TMT HROS**

  Two feasibility studies were undertaken for a high-resolution spectrograph (HROS) for TMT. In contrast to CODEX, the main priority is sensitivity rather than stability, so HROS would be mounted at one of the Nasmyth foci. The HROS-MTHR concept has blue and red channels, each of which is a cross-dispersed echelle spectrograph (Figure 5). This relatively 'classic' design results in a large space-envelope for the instrument (~11x10x4 m) and large optical elements, e.g., 1.6 m off-axis collimators, 1.4 m camera lenses, and a huge echelle mosaic. The CU-HROS concept took a very different approach, splitting the spectral range with dichroics into 32 spectral bins, which are then relayed to 32 separate echelle spectrographs [16].

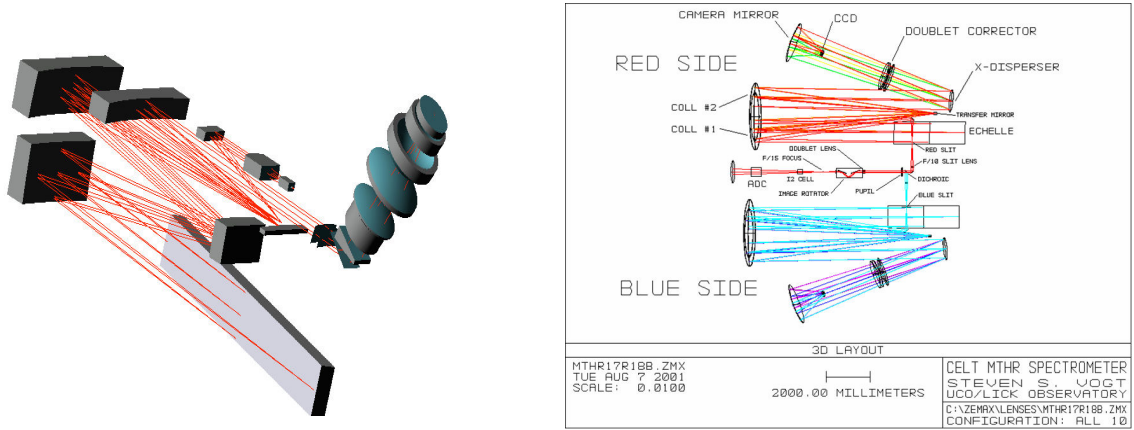

Figure 5: Optical design of the CODEX spectrographs [left, 17]; HROS-MTHR concept for TMT [right, 18].

# SEEING-LIMITED, INTERMEDIATE-RESOLUTION SPECTROSCOPY

The Wide-Field Optical Spectograph (WFOS) is a first-light TMT instrument for $R \sim 5,000$, multi-object spectroscopy at visible wavelengths. The highlight science case is tomographic mapping of the intergalactic medium via observations of background galaxies. With current facilities we are limited to bright quasars, i.e. relatively sparse sampling, whereas TMT will revolutionize the range of accessible sight-lines accessible as fainter galaxies will come into reach. The main performance drivers for WFOS are for the largest possible multiplex (>1,000), over the largest possible field-of-view. The first WFOS design presented to the community [19] comprised four 'barrels' (as shown in Figure 6) that were, in effect, separate instruments. This has now been scaled-back to two channels, each with a field-of-view of 4.5'x5.4' which, with some slight vignetting gives a total field of 46 arcmin$^2$. Note that the over-sampling discussed earlier is a feature of the design, with typical slit-widths of ~0.75" sampled at 0.07"/pixel.

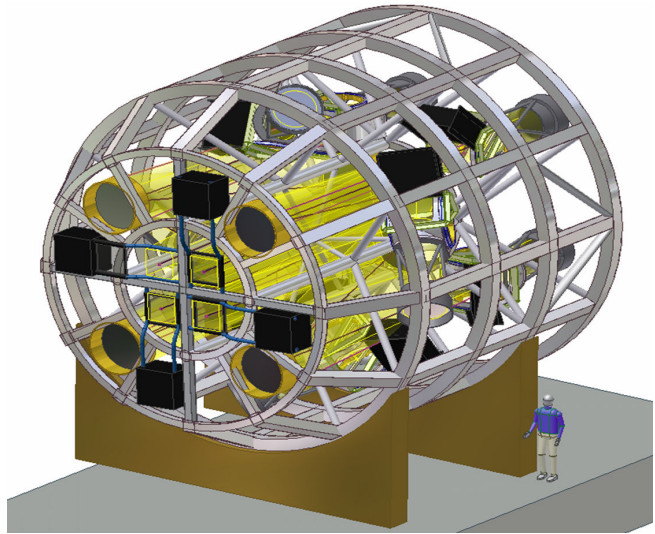

**Figure 6: TMT-WFOS four-channel conceptual design [19].**

The case can be made for a similar instrument on the E-ELT, viewing it as a 'workhorse' multi-object spectrograph in the same vein as VLT-FORS, or Gemini-GMOS, although the problems of over-sampling become even more acute at the plate-scale of the E-ELT. A technical solution for target acquisition could be a fibre-fed instrument, e.g. a large OzPoz-like fibre positioner [20] that deploys fibres that relay to batteries of '8-m class' spectrographs (*ala* VLT-MUSE) or, in the long-term future, perhaps even miniature photonic spectrographs. The caveat to this is that if a similar multiplex to WFOS is sought, such an instrument would be even larger and more expensive for the 42-m E-ELT design. It is timely to consider such a study for the E-ELT, although this will obviously depend on the formation of a suitably motivated and competent consortium. Moreover, if adopted as a first-generation instrument, one of the existing studies will likely have to be shelved owing to budgetary constraints.

While the majority of the ELT cases at visible wavelengths can be satisfied by seeing-limited observations, there remains a number of compelling cases for high-stability, diffraction-limited performance in the visible. Even considering technologies currently under development (i.e. with low readiness levels), such performance is very unlikely from a ground-based facility for the foreseeable future and we might have to look to new space-based instrumentation for the solution. For instance, one of the long-term development goals from the *James Webb Space Telescope (JWST)* has been for an efficient deployment method that could scale to yet larger space telescope. Meanwhile a conceptual design has recently been undertaken for an 8-m monolithic space telescope. Both of these options are clearly long-term solutions!

## CONCLUSIONS

It is clear that there is no such thing as a 'simple' instrument for an ELT, and even first-light instruments (e.g. TMT-WFOS) will be very challenging projects. The science cases advanced for ELTs cover the whole range of spatial resolutions, from the diffraction limit up to seeing-limited observations – these will require very different types of instruments, each with their difficulties. In terms of both size and stability, instruments for the ELTs will likely be an order of magnitude more demanding than those on 8 and 10-class facilities, not to mention the fact that the incorporation of on-instrument AO systems will become increasingly common. Indeed, the AO simulations presented here illustrate the potential for excellent spatial-resolution in the near-infrared. Studies are now well-advanced on both sides of the Atlantic to exploit this new parameter space, both for diffraction-limited observations, and for multi-object spectroscopy over wider-fields.

Finally, we note that a large number of the science cases advanced for ELTs require some level of information at optical wavelengths, e.g. the assembled examples in the OPTICON E-ELT Science Case [21]. To some extent this probably reflects the long history of optical astronomy as compared to infrared observations, but it is also true that in some areas of research the most useful diagnostic features (be they spectral lines or photometric colours) are at visible wavelengths. As such, optical observations will prove an attractive prospect for large sections of the community and, while the instruments may be both large and expensive, they are certainly feasible.

*Acknowledgements:* We thank David Crampton for kindly providing the TMT information and images, and Mark Casali for the image of the wide-field GLAO concept.

**REFERENCES**

1. Hubin, N. et al., in *The Scientific Requirements for Extremely Large Telescopes*, P. Whitelock, M. Dennefeld and B. Leibundgut, eds., IAU Symposium No. 232, Cambridge University Press, 2006, p60
2. Lorente, N. P. F. et al., 2006, Proc. SPIE, 6274, 44
3. Lenzen, R. et al., 2003, Proc. SPIE, 4841, 944
4. Dohlen, J. et al., 2006, Proc. SPIE, 6269, 24
5. Vérinaud, C. et al., 2006, Proc. SPIE, 6272, 19
6. Eisenhauer, F. et al., 2003, Proc. SPIE, 4841, 1548
7. McLean, I. S. & Adkins, S. M., 2006, Proc. SPIE, 6269, 1
8. Herriot, G. et al., 2006, Proc. SPIE, 6272, 22
9. Tecza, M. et al., to appear in *Science with the VLT in the ELT era*, A. Moorwood, ed., Springer Astrophysics and Space Science Proceedings
10. Genzel, R. et al., 2006, Nature, 442, 786
11. Le Louran, M. & Hubin, N., 2004, MNRAS, 349, 1009
12. Fowler, A. et al., 2003, Proc. SPIE, 4841, 853
13. Probst, R. et al., 2004, Proc. SPIE, 5492, 1716
14. Jackson, J. C. et al., 2003, Opt. Eng. 42, 112
15. Shearer, A. et al., these proceedings
16. Froning, C. et al., 2006, Proc. SPIE, 6269, 61
17. Pasquini, L. et al., in *The Scientific Requirements for Extremely Large Telescopes*, P. Whitelock, M. Dennefeld and B. Leibundgut, eds., IAU Symposium No. 232, Cambridge University Press, 2006, p193
18. Crampton, D. & Simard, L., 2006, Proc. SPIE, 6269, 59
19. Padzer et al., 2006, Proc. SPIE, 6269, 63
20. Gillingham, P. et al., 2003, Proc. SPIE, 4841, 1170
21. Hook, I. ed., *The Science Case for the European Extremely Large Telescope,* OPTICON, 2005